# Influence of Surface Functionalization on the Colloidal Stability and Magnetic Properties of Ferrite Nanoparticles

Shaikh Arif Shaikh Iqbal[a,b], Radha Srinivasan[a]

[a] Department of Physics, University of Mumbai, Maharashtra, India
[b] Anjuman-I-Islam's Kalsekar Technical Campus, School of Engineering & Technology, New Panvel, Maharashtra, India

**Abstract:**

Magnetite ($Fe_3O_4$) nanoparticles have attracted considerable interest due to their tunable physicochemical properties and relevance in functional materials. In this work, uncoated $Fe_3O_4$ nanoparticles and surface-modified $Fe_3O_4$ nanoparticles coated with polyethylene glycol (PEG-6000) and citric acid (CA) were synthesized via a chemical co-precipitation method. The effect of surface functionalization on the structural, spectroscopic, magnetic, and colloidal properties of the nanoparticles was systematically investigated. X-ray diffraction analysis confirmed the formation of phase-pure $Fe_3O_4$ with an inverse spinel structure for all the samples. Fourier transform infrared and Raman spectroscopy verified successful surface modification while preserving the Fe-O framework of the magnetite core. Dynamic light scattering and zeta potential measurements indicated improved dispersion and colloidal stability for the surface-modified nanoparticles. Magnetization studies performed at room temperature revealed superparamagnetic behaviour for all samples, accompanied by a coating-dependent reduction in saturation magnetization. Overall, the results emphasize the critical role of surface chemistry in tailoring the physicochemical, magnetic, and colloidal behaviour of $Fe_3O_4$ nanoparticles.

## 1. Introduction

Magnetic iron oxide nanoparticles, particularly magnetite ($Fe_3O_4$), have been widely studied due to their unique size-dependent magnetic behaviour, chemical stability, and versatility in functional material design [1-3]. At the nanoscale, $Fe_3O_4$ exhibits properties such as superparamagnetism, high saturation magnetization, and relatively low toxicity, making it suitable for applications in catalysis, sensing, environmental remediation, and biomedicine [4-6]. Among the various iron oxide phases, $Fe_3O_4$ is especially favoured because of its inverse spinel structure and strong magnetic response [7].

Several synthetic routes have been reported for the preparation of $Fe_3O_4$ nanoparticles, including thermal decomposition, sol-gel processing, hydrothermal synthesis, and chemical co-precipitation [8–10]. Among these, co-precipitation remains one of the most widely followed methods due to its simplicity, scalability, cost-effectiveness, and ability to produce nanoparticles under relatively mild conditions [11]. However, $Fe_3O_4$ nanoparticles synthesized via aqueous co-precipitation often suffer from agglomeration caused by strong magnetic dipole-dipole interactions and high surface energy, that adversely affect particle size distribution and colloidal stability [12].

Surface functionalization is an effective strategy to overcome these limitations by modifying the nanoparticle interface without altering the intrinsic crystal structure of the magnetic core [13,14]. Appropriate surface coatings can restrict particle growth, enhance dispersion stability, and tailor interfacial charge. Polyethylene glycol (PEG) is one of the most commonly employed surface modifiers owing to its hydrophilicity, chemical stability, and ability to provide steric stabilization, resulting in improved dispersion and reduced nonspecific interactions [15,16]. Similarly, citric acid also widely used as a surface-modifying agent due to its multidentate carboxylate groups, which strongly coordinate with surface Fe ions and impart electrostatic stabilization [17,18].

Although many studies have reported PEG or citric-acid-coated $Fe_3O_4$ nanoparticles, systematic comparative investigations in which uncoated and surface-modified nanoparticles are synthesized under identical conditions remain limited [19-21]. In this context, the present work reports the synthesis of uncoated $Fe_3O_4$ nanoparticles and $Fe_3O_4$ nanoparticles surface modified with PEG-6000 and citric acid via a co-precipitation route, followed by comprehensive physicochemical characterization.

In addition to structural and surface properties, the magnetic behaviour of $Fe_3O_4$ nanoparticles plays an important role in determining their functional performance. Surface modification can influence magnetic parameters such as saturation magnetization and coercivity due to surface spin disorder and mass dilution effects. Therefore, a combined assessment of physicochemical, colloidal, and magnetic properties is essential to understand the impact of surface functionalization on $Fe_3O_4$ nanoparticles [22,23].

## 2. Materials and Methods

### 2.1 Materials

Ferric chloride ($FeCl_3 \cdot 6H_2O$) and ferrous chloride ($FeCl_2 \cdot 4H_2O$) were used as iron precursors. Aqueous ammonia solution (25%, w/w) was employed as the precipitating agent. Polyethylene glycol (PEG-6000) and citric acid were used as surface modifying agents. All chemicals were of analytical grade and used as received. Deionized water was used throughout the synthesis, and ethanol was used during washing and dispersion steps.

### 2.2 Synthesis of $Fe_3O_4$ Nanoparticles (A1)

Uncoated $Fe_3O_4$ nanoparticles (A1) were synthesized using a conventional chemical co-precipitation method. $FeCl_3 \cdot 6H_2O$ (10.812 g) and anhydrous $FeCl_2$ (2.535 g) were dissolved in 200 mL water, maintaining a $Fe^{3+}:Fe^{2+}$ molar ratio of 2:1 under continuous magnetic stirring. The reaction mixture was heated to 65 °C under ambient air atmosphere, and aqueous ammonia was added dropwise until the pH reached between 12 to 14, resulting in the formation of a black precipitate characteristic of $Fe_3O_4$ nanoparticles. After completion of precipitation, the nanoparticles were separated by magnetic decantation and washed repeatedly with deionized water and ethanol until neutral pH was achieved. After supernatant removal the nanoparticles were dried in hot air oven at 50°C for 10 hours. This synthesis protocol follows a widely reported co-precipitation approach for magnetite nanoparticles synthesis [11,22].

## 2.3 Surface Functionalization of Fe$_3$O$_4$ Nanoparticles

### 2.3.1 PEG-6000 Coating (A2)

PEG-6000-coated Fe$_3$O$_4$ nanoparticles (A2) were prepared via a post-synthesis surface functionalization approach. After washing and pH neutralization, the Fe$_3$O$_4$ nanoparticle suspension was reheated to 65 °C under continuous stirring. Subsequently, 0.5 g of PEG-6000 was added, and the suspension was stirred for 1 hr to allow effective adsorption of PEG chains onto the nanoparticle surface. The coated nanoparticles were magnetically separated and repeatedly washed with deionized water and ethanol to remove excess amount of polymer. The drying procedure is repeated for coated nanoparticles as of uncoated nanoparticles.

### 2.3.2 Citric Acid Coating (A3)

Citric-acid-coated Fe$_3$O$_4$ nanoparticles (A3) were prepared by introducing citric acid post synthesis following the same procedure as of A2 preparation. Citric acid molecules coordinate with surface Fe ions through carboxylate groups, resulting in electrostatic stabilization of the nanoparticles. The product was separated magnetically and washed thoroughly prior to characterization. Surface modification using polyethylene glycol and citric acid was carried out following established approaches reported for improving dispersion and stability of iron oxide nanoparticles [15,18].

## 2.4 Characterization Techniques

The crystal structure and phase purity of the synthesized nanoparticles were examined using powder X-ray diffraction (XRD) with a Bruker D8 Advance diffractometer employing Cu Kα radiation (λ = 1.5418 Å) in a coupled θ-2θ configuration.

Fourier transform infrared (FTIR) spectroscopy was performed using a Bruker Vertex 80 FTIR spectrometer coupled with a Hyperion 3000 microscope to identify surface functional groups. Raman spectra were recorded using a Renishaw Raman spectrometer (Model: INCIA0120-20) to analyze vibrational characteristics and phase nature.

The hydrodynamic particle size distribution was measured by dynamic light scattering (DLS), and the surface charge was evaluated by zeta potential measurements using a Malvern Zetasizer Nano ZS90. All measurements were carried out at room temperature using ethanol as the dispersant.

The magnetic properties of the nanoparticles were investigated using a Quantum Design vibrating Sample Measurement at 300 K. Magnetization as a function of applied magnetic field was recorded to evaluate saturation magnetization, coercivity, and remanent magnetization.

## 3. Results and Discussion

### 3.1 Structural Analysis by X-ray Diffraction (XRD)

The X-ray diffraction patterns of uncoated Fe$_3$O$_4$ nanoparticles (A1), PEG-6000–coated Fe$_3$O$_4$ nanoparticles (A2), and citric-acid-coated Fe$_3$O$_4$ nanoparticles (A3) are shown in Figure 1. All samples exhibit diffraction peaks corresponding to the characteristic reflections of inverse spinel magnetite (Fe$_3$O$_4$), confirming successful synthesis of the desired phase.

The prominent diffraction peaks observed at 30.2°, 35.6°, 43.3°, 53.6°, 57.3°, and 62.9° can be indexed to the (220), (311), (400), (422), (511), and (440) crystallographic planes of $Fe_3O_4$, respectively. These reflections are in good agreement with standard data reported for magnetite nanoparticles synthesized via co-precipitation methods [7,11]. No additional peaks corresponding to secondary iron oxide phases such as hematite ($\alpha$-$Fe_2O_3$) or maghemite ($\gamma$-$Fe_2O_3$) were detected, indicating high phase purity.

The diffraction peaks for all samples exhibit noticeable broadening, confirming the nanocrystalline nature of the particles. The calculated average crystallite sizes were found to be 26.0 nm for A1, 18.7 nm for A2, and 18.4 nm for A3. The average crystallite size, estimated using the Scherrer equation, shows a clear reduction for surface-modified samples compared to uncoated $Fe_3O_4$. This reduction in crystallite size for PEG-6000 and citric-acid-coated nanoparticles can be attributed to restricted crystal growth during nucleation due to steric hindrance and surface coordination effects, as commonly reported for surface-functionalized magnetite systems [22]. Importantly, the absence of peak shifts across all samples suggests that surface modification does not alter the intrinsic crystal structure of the $Fe_3O_4$ core.

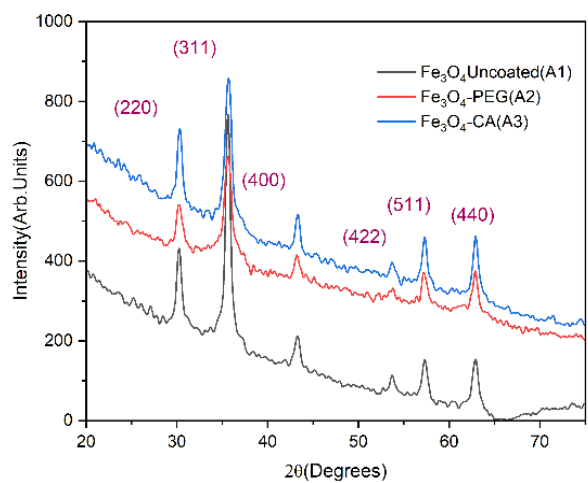

Fig.1: XRD patterns of uncoated and coated iron oxide nanoparticles.

### 3.2 FTIR Analysis

Figure 2 presents the FTIR spectra of uncoated and surface-modified $Fe_3O_4$ nanoparticles. All samples exhibit a strong absorption band in the low-wavenumber region (~570-630 $cm^{-1}$), which is attributed to the Fe-O stretching vibration of magnetite, confirming preservation of the iron oxide core structure [15].

The spectrum of uncoated $Fe_3O_4$ (A1) shows a broad band around ~3400 $cm^{-1}$, corresponding to O-H stretching vibrations arising from surface-adsorbed water molecules and hydroxyl groups. In the case of PEG-6000-coated $Fe_3O_4$ (A2), additional characteristic bands appear near ~2920 and ~2850 $cm^{-1}$, corresponding to asymmetric and symmetric stretching vibrations of -$CH_2$ groups, along with a band around ~1090 $cm^{-1}$ attributed to C-O-C stretching vibrations of PEG chains. These features confirm successful PEG-6000 functionalization, consistent with previously reported PEG-coated iron oxide nanoparticles [16].

For citric-acid-coated $Fe_3O_4$ nanoparticles (A3), distinct absorption bands are observed near ~1620 cm$^{-1}$ and ~1400 cm$^{-1}$, corresponding to asymmetric and symmetric stretching vibrations of carboxylate (-COO$^-$) groups. The presence of these bands indicates coordination of citric acid molecules to surface iron ions, which is in agreement with established reports on citric-acid-functionalized magnetite nanoparticles [17,18]. The FTIR results clearly shows surface modification was successfully achieved without disrupting the $Fe_3O_4$ lattice.

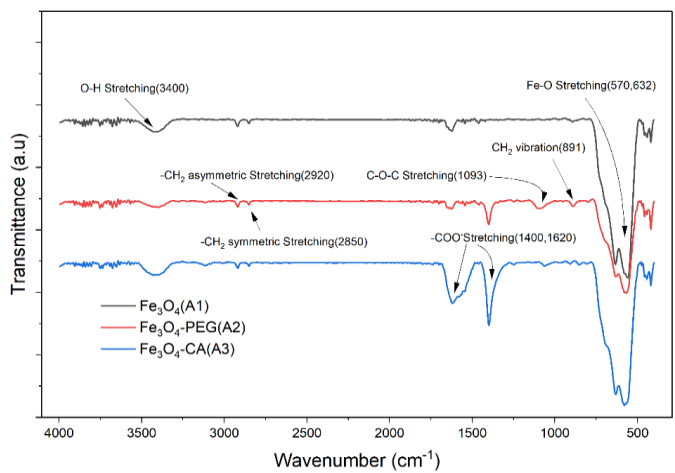

Fig. 2: FTIR spectrum of uncoated(A1) and PEG coated (A2) and Citric acid coated Fe3O4 nanoparticles(A3).

### 3.3 Raman Spectroscopy

Raman spectra of the synthesized nanoparticles are shown in Figure 3. The uncoated $Fe_3O_4$ nanoparticles (A1) exhibit characteristic Raman bands at 670 cm$^{-1}$, which are associated with the $T_2g$ and $A_1g$ vibrational modes of magnetite. These Raman features are consistent with reported vibrational signatures of $Fe_3O_4$ nanoparticles [7,9].

For PEG-6000-coated $Fe_3O_4$ nanoparticles (A2), the dominant Fe-O vibrational band appears slightly shifted and broadened compared to A1, indicating surface-induced strain and size effects resulting from polymer coating. Citric-acid-coated $Fe_3O_4$ nanoparticles (A3) show a similar Fe-O-related band with further broadening and minor shifts, which can be attributed to strong surface coordination between carboxylate groups and iron ions. Such coating-induced Raman peak modifications have been previously observed in surface-functionalized magnetite systems and are commonly associated with reduced crystallite size and surface disorder [22].

Overall, the Raman results support the XRD findings and confirm that surface modification influences interfacial vibrational characteristics without causing phase transformation. Weak and broad features observed in the 1300–1500 cm$^{-1}$ region for all samples are attributed to second-order Raman scattering (two-phonon processes) associated with strong Fe-O vibrational modes of magnetite, with possible minor contributions from surface-bound organic groups in coated nanoparticles.

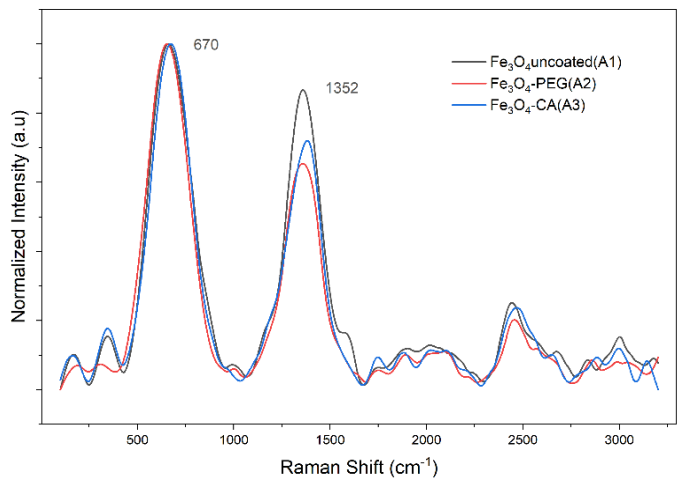

Fig.3: Raman Spectra of uncoated and coated iron oxide nanoparticles.

### 3.4 Particle Size Analysis

The hydrodynamic particle size distributions of A1, A2, and A3 measured by DLS are presented in Figure 4. All samples exhibit hydrodynamic diameters in the nanometer range (99nm for A1, 90nm for A2 and 96nm for A3), with values significantly larger than the crystallite sizes obtained from XRD. This difference is expected, as DLS measures the effective hydrodynamic diameter, which includes the particle core, surface coating, solvation layer, and any weakly associated aggregates [12].

Compared to uncoated $Fe_3O_4$ (A1), both PEG-6000 and citric acid coated nanoparticles show improved dispersion behaviour, reflected by relatively narrower size distributions. The reduced agglomeration observed for coated samples can be attributed to steric stabilization by PEG-6000 and electrostatic repulsion arising from surface-bound carboxylate groups in citric acid-modified nanoparticles, as reported in earlier studies [16,18].

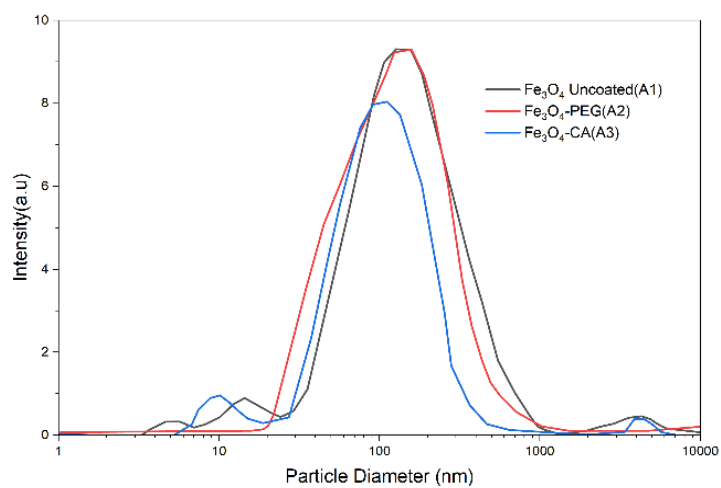

Fig.4: Hydrodynamic particle size of uncoated and coated iron oxide nanoparticles.

## 3.5 Zeta Potential Analysis

Zeta potential measurements were performed in ethanol as the dispersing medium to evaluate the surface charge behaviour of the synthesized nanoparticles. As shown in figure 5 all samples exhibited positive zeta potential values, indicating that the nanoparticle surfaces remain positively charged under the measurement conditions. It is well known that the surface charge of magnetite nanoparticles is highly sensitive to the surrounding environment, as surface hydroxyl groups can undergo protonation or deprotonation depending on the medium and surface chemistry [17]. In non-aqueous solvents such as ethanol, the dissociation of surface functional groups is reduced, which can result in positively charged surfaces even after surface modification.

Surface coatings are also known to influence the position of the slipping plane and, consequently, the measured zeta potential without altering the magnetite core [22]. In the present study, PEG-6000-coated $Fe_3O_4$ nanoparticles exhibited the highest positive zeta potential, which can be attributed to steric effects and solvent-polymer interactions in ethanol. Citric acid coated $Fe_3O_4$ nanoparticles showed moderately positive zeta potential values, suggesting partial ionization of surface carboxyl groups in the non-aqueous medium. Overall, these results highlight the solvent and coating-dependent nature of zeta potential in iron oxide nanoparticle systems and confirm that positive zeta potential values are reasonable for $Fe_3O_4$ nanoparticles dispersed in ethanol.

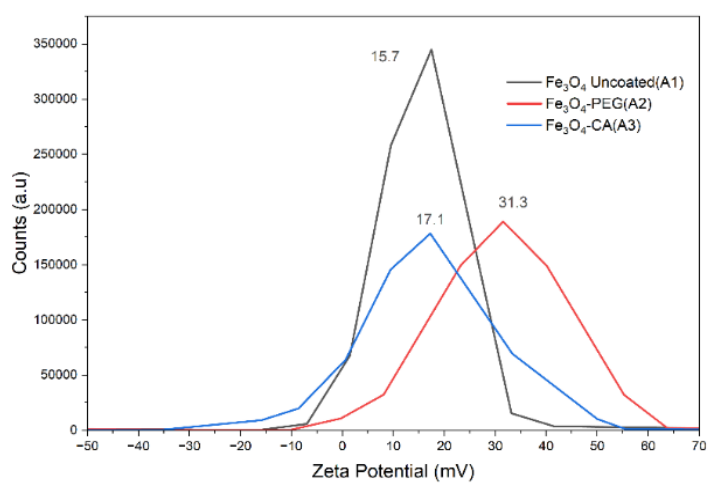

Fig.5: Zeta potential of uncoated and coated iron oxide nanoparticles.

## 3.6 Magnetic Properties Analysis

The magnetic properties of A1, A2 and A3 were investigated using Quantum Design PPMS at 300K. The magnetization as a function of applied magnetic field is shown in figure 6, and the extracted magnetic parameters are summarized in Table 1.

All samples exhibit S-shaped M-H curves with negligible coercivity and remanent magnetization, confirming their superparamagnetic behaviour at room temperature. The saturation magnetization (Ms) of uncoated $Fe_3O_4$ nanoparticles (71.3 emu/g) is higher than that of surface modified samples. A systematic decrease in Ms is observed upon surface functionalization, with PEG-6000 coated $Fe_3O_4$ showing the lowest Ms (57.5 emu/g), followed by citric-acid-coated $Fe_3O_4$ (64.9 emu/g).

The reduction in Ms after surface modification can be attributed to the presence of a non-magnetic organic shell and enhanced surface spin disorder, which reduce the effective magnetic contribution per unit mass. Importantly, the near-zero coercivity (Hc ≈ 0-15 Oe) and low remanent magnetization for all samples indicate the absence of magnetic hysteresis, demonstrating that surface functionalization does not compromise the superparamagnetic nature of $Fe_3O_4$ nanoparticles [23].

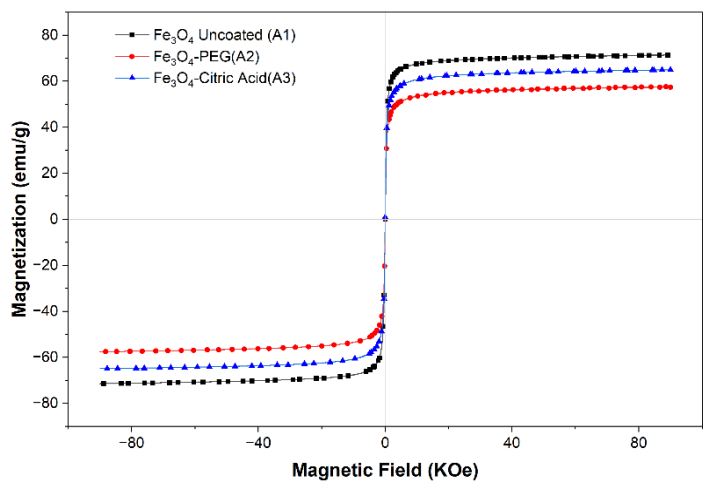

Fig.6: Magnetization curve of uncoated and coated Ferrite Nanoparticles.

| Sample | Coating | XRD size(nm) | DLS size(nm) | Zeta (mV) | Ms(emu/g) | Hc (Oe) | Mr (emu/g) |
|---|---|---|---|---|---|---|---|
| A1 | --- | 26 | 99 | 15.7 | 71.3 | ~0 | ~0.9 |
| A2 | PEG6000 | 19 | 91 | 31.3 | 57.5 | ~0 | ~0.4 |
| A3 | Citric Acid | 18 | 97 | 17.1 | 64.9 | ~15 | ~2.1 |

Table 1: Summary of Size, Surface Charge, and Magnetic Parameters of Ferrite Nanoparticles

### 3.7 Comparative Assessment

The comparative evaluation of uncoated $Fe_3O_4$ (A1), PEG-6000-coated $Fe_3O_4$ (A2), and citric acid coated $Fe_3O_4$ (A3) nanoparticles highlights the decisive role of surface functionalization in tuning physicochemical, colloidal, and magnetic properties while preserving the intrinsic crystal structure of magnetite.

Structural and spectroscopic analyses confirm that all samples retain the inverse spinel phase of $Fe_3O_4$, indicating that surface modification is confined to the nanoparticle interface. The reduced crystallite size observed for surface-modified samples suggests restricted crystal growth during synthesis, which can be attributed to steric hindrance in the presence of PEG-6000 and surface coordination effects associated with citric acid, in line with earlier studies [13].

Surface functionalization has a pronounced influence on dispersion behaviour. For all samples, hydrodynamic particle sizes are larger than the corresponding crystallite sizes due to solvation effects and interfacial contributions. However, both PEG-6000 and citric-acid-coated nanoparticles exhibit reduced agglomeration compared to uncoated $Fe_3O_4$. PEG-6000 provides effective stabilization primarily through steric interactions, whereas citric acid improves dispersion through strong surface coordination and modification of interparticle interactions [16,18].

Zeta potential measurements performed in ethanol further support the role of surface chemistry in governing colloidal behaviour. All samples display positive zeta potential values, reflecting solvent-dependent surface charging of iron oxide nanoparticles. The higher positive zeta potential observed for PEG6000 coated $Fe_3O_4$ indicates enhanced stability associated with polymer-mediated steric effects and changes in the slipping plane, while citric acid coated nanoparticles show moderately positive values consistent with partial surface charge modification in the non-aqueous medium [17]. Overall, PEG6000 emerges as the more effective stabilizing agent under the present conditions, while citric acid offers strong surface interaction without altering the $Fe_3O_4$ core structure.

Magnetization measurements indicate that all samples retain superparamagnetic behaviour at room temperature. The observed decrease in saturation magnetization upon surface functionalization is consistent with coating-dependent magnetic behaviour reported for $Fe_3O_4$ nanoparticles and can be attributed to non-magnetic surface layers and surface spin disorder effects [23].

## 4. Conclusion

Uncoated and surface-modified $Fe_3O_4$ nanoparticles were successfully synthesized by a co-precipitation method, yielding phase-pure magnetite with an inverse spinel structure. Surface functionalization with PEG6000 and citric acid was achieved without altering the crystal structure of $Fe_3O_4$ core, confirming the robustness of the synthesis approach.

The results demonstrate that surface chemistry plays a key role in controlling particle dispersion, interfacial behaviour, and magnetic response. In particular, PEG6000 coating provided improved colloidal stability under the studied conditions, while citric acid enabled effective surface coordination. These findings establish a reliable physicochemical and magnetic baseline for $Fe_3O_4$ nanoparticles and support their further investigation in advanced functional and biological studies.

**Acknowledgements**

The authors thank the Head and staff of the Department of Physics, University of Mumbai, for their timely support. The authors also acknowledge the UGC-DAE Consortium for Scientific Research (CSR), Mumbai Centre, for providing magnetization measurement facilities and thank the Director and faculty members for their assistance. The support of the Director and Dean, School of Pharmacy, A.I Kalsekar Technical Campus (AIKTC), New Panvel, is also gratefully acknowledged.

The authors further acknowledge the instrumentation facilities provided by SAIF, IIT Bombay, and CFC-SAIF, Shivaji University, Kolhapur, accessed through the iSTEM facility, for characterization support.

**Declaration of generative AI and AI-assisted technologies in the manuscript preparation process**

During the preparation of this work the author used Chatgpt-5.1 for checking sentence formation and making it grammatically acceptable. After using this tool/service, the author/co-author reviewed and edited the content as needed and take full responsibility for the content of the published article.